\documentclass[12pt]{iopart}
\usepackage{graphicx,psfig,epsfig}
\def\la{\langle}
\def\ra{\rangle}
\def\beq{\begin{equation}}
\def\eeq{\end{equation}}
\def\be{\begin{eqnarray}}
\def\ee{\end{eqnarray}}

\def\k2av{\la k_T^2\ra}
\newcommand{\f}[2]{\frac{#1}{#2}}                                           
\newcommand{\dd}{ {\textrm d}}              

\begin{document}

\title[Cronin Effect at Different Rapidities at RHIC]{Cronin 
Effect at Different Rapidities at RHIC}

\author{G G Barnaf\"oldi\dag\ddag , G Papp\ddag , P L\'evai\dag , G Fai\P }

\address{\dag\ RMKI KFKI, P.O. Box 49, Budapest, H-1525, Hungary} 
\address{\ddag\ E\"otv\"os University, P\'azm\'any P. 1/A, 
Budapest, H-1117, Hungary}
\address{\P\ CNR, Kent State University, Kent, OH 44242, USA} 

\ead{bgergely@rmki.kfki.hu}

\begin{abstract}
Calculations of the nuclear modification factor, $R_{dAu}$, 
for $\pi^0$ production in $dAu$ collisions at 
${\sqrt s_{NN}} = 200$ GeV
are presented. The applied pQCD-improved parton model incorporates
intrinsic $k_T$. Nuclear multiscattering and nuclear shadowing are
considered in the $Au$ nucleus. Theoretical results are displayed
for midrapidity and high pseudorapidity, $\eta $, 
and compared to preliminary PHENIX and BRAHMS data.     
\end{abstract}
\submitto{\JPG}


\section*{Introduction}
Midrapidity $AuAu$ experiments at RHIC energies display strong suppression in  
$\pi$ spectra at high $p_T$~\cite{phenix_QM01}. This effect 
was explained by jet energy loss
\mbox{in hot dense matter} (see Ref.~\cite{LP_QM01}). To investigate whether 
initial state effects play a role in this suppression, $dAu$ 
experiments were carried out. Instead of suppression, an enhancement has
been seen in minimum bias data \cite{bias_phenix} (the ``Cronin 
effect'' in $pA$ collisions at FERMILAB energies \cite{Cron75,Antr79}). 
This result indicates the absence of extra
shadowing (gluon saturation), related to proposed wave-function 
modifications in the fast moving $Au$ nucleus (see Ref.~\cite{CGC}). 
However, new data with centrality
dependence in midrapidity $dAu$ collisions \cite{pre_phenix} could 
provide more detailed information about the interplay between 
nuclear multiscattering \cite{PG_QM01} and (standard) nuclear shadowing 
\cite{Shadxnw_uj}. Analysis of recent data at forward 
rapidities~\cite{pre_brahms} can give more
insight into the $\eta$-dependence of nuclear effects.

Here we display the results of our NLO pQCD calculations on
pion production at \mbox{$2 < p_T < 10$ GeV} at different centralities.
In parallel, we consider available data at forward
rapidities in $dAu$ collisions \cite{pre_brahms}, and investigate them
theoretically.

\section{Calculational method in a nutshell}

We perform calculations on $\pi^0$ production in $dAu$ collisions using 
a pQCD-improved parton model extended by a Glauber-type 
collision geometry~\cite{pgNLO,Aversa89,Aur00}. Introducing 
nuclear thickness function, $t_{Au}$, with
a Woods\,--\,Saxon formula for $Au$ and $t_d$ with a sharp sphere
for $d$, the invariant cross section 
is obtained as~\cite{dAu}:
\begin{eqnarray}
\label{hadX}
& E_{\pi} & \f{\dd \sigma_{\pi}^{dAu}}{ \dd ^3p} =
             \int \dd ^2b \, \dd ^2r 
            \, \, t_d(r) \, \, t_{Au}(\vert{\bf b} - {\bf r} \vert) \, \,
            \f{1}{s} \, \sum_{abc}  
            \int^{1-(1-v)/z_c}_{vw/z_c}\! \f{\dd \hat{v}}{\hat{v}(1-\hat{v})}\!
            \times \nonumber \\
& \times &  \int^{1}_{vw/\hat{v}z_c} \f{ \dd \hat{w} }{\hat{w}} \!
	    \int^1 {\dd z_c} 
            \int \!\! {\dd^2 {\bf k}_{Ta}} \!\! \int \!\! {\dd^2 {\bf k}_{Tb}}
            \,\, f_{a/d}(x_a,{\bf k}_{Ta},Q^2)
            \,\, f_{b/A}(x_b,{\bf k}_{Tb},Q^2) \times \nonumber \\ 
& \times &  \left[
            \f{\dd {\widehat \sigma}}{\dd \hat{v}} \delta (1-\hat{w})\, + \,
            \f{\alpha_s(Q_r)}{ \pi}  K_{ab,c}(\hat{s},\hat{v},\hat{w},Q,Q_r,\tilde{Q}) \right] 
            \f{D_{c}^{\pi} (z_c, \tilde{Q}^2)}{\pi z_c^2}  \,\,  .
\end{eqnarray}
In our next-to-leading order (NLO) calculation~\cite{pgNLO},                
$\dd {\widehat \sigma}/ \dd \hat{v}$ represents the Born cross section      
of the partonic subprocess, and                                              
$K_{ab,c}(\hat{s},\hat{v},\hat{w},Q,Q_r,\tilde{Q})$ is the corresponding    
higher order correction term~\cite{pgNLO,Aversa89,Aur00}.        
We fix the factorization and renormalization                                
scales and connect them to the momentum of the intermediate jet,            
$Q=Q_r=(4/3) p_q$ (where $p_q=p_T/z_c$), reproducing $pp$ data at RHIC     
with high precision at high $p_T$ \cite{dAu}.                               
                                                                            
The approximate form of the 3-dimensional parton distribution 
function (PDF) is:                                                           
\begin{equation}                                                            
f_{a/p}(x_a,{\bf k}_{Ta},Q^2) \,\,\,\, = \,\,\,\, f_{a/p}(x_a,Q^2)          
\cdot g_{a/p} ({\bf k}_{Ta}) \ .                                            
\end{equation}                                                              
Here, the function $f_{a/p}(x_a,Q^2)$ represents the standard NLO 
PDF as a function of momentum fraction of the incoming parton, $x_a$, at     
scale $Q$ (we use MRST(cg)). The
partonic transverse-momentum distribution is defined phenomenologically as a 
2-dimensional Gaussian, $g_{a/p}({\bf k}_T)$, and characterized by an 
``intrinsic $k_T$'' width~\cite{pgNLO,YZ02}.            
                                                                            
Nuclear multiscattering is accounted for through 
broadening of the incoming parton's $k_T$-distribution function,
namely an increase in the width of the Gaussian:  
\beq                                                                        
\label{ktbroadpA}                                                           
\k2av_{pA} = \k2av_{pp} + \Delta(b) \,\,\,\,\,\,\,\, \textrm{where} \,\,\,\,\,\,\,  \Delta (b) = C \cdot h_{pA}(b) \ .                             
\eeq      
Pion production in $pp$ collisions at RHIC energy indicates
the value \mbox{$\k2av_{pp}=2.5$ GeV$^2$} \cite{dAu,YZ02}.
Considering the multiscattering part, 
$h_{pA}(b)$  describes the number of {\it effective} $NN$ collisions 
(or {\it effective} collision length for partons) at impact 
parameter $b$, which impart an average transverse momentum squared, 
$C$~\cite{YZ02}.
At SPS energies
the effectivity function $h_{pA}(b)$ was written in terms of the   
number of collisions suffered by the incoming proton in the target nucleus. 
For this energy region we have found  a limited number of semihard collisions, 
$\nu_{m} -1 = 4$ and the value $C = $ 0.35 GeV$^2$~\cite{Bp02}, 
which implies a total broadening $\Delta \sim 1$ GeV$^2$. 
We will assume these values at RHIC energies, although the
precise energy and
rapidity dependence of $\nu_m$ and $C$ is not yet verified.

The nuclear PDF, $f_{b/A}(x_b,{\bf k}_{Tb},Q^2)$, incorporates the 
shadowing parametrization used in HIJING~\cite{Shadxnw_uj}.
Extra gluon saturation (CGC)~\cite{CGC} would require a further
suppression factor to be introduced. The experimental data
will decide if such an extra suppression factor is necessary.

Fragmentation function (NLO KKP), $D_{c}^{\pi}(z_c, \tilde{Q}^2)$, 
in eq.~(\ref{hadX}) is responsible for the hadronization                    
of parton $c$ into a pion with momentum        
fraction $z_c$ at scale $\tilde{Q}=(4/3) p_T$.    

We present results on the nuclear modification factor, defined as follows:
\begin{equation}
R_{dAu}= \frac{1}{N_{bin}}
\frac{E_{\pi}\dd \sigma^{dAu}_{\pi}/ \dd ^3 p}{E_{\pi}\dd \sigma^{pp}_{\pi}/ \dd ^3 p}
=\frac{E_{\pi}\dd \sigma^{dAu}_{\pi}(\textrm{with nuclear effects})/ \dd ^3 p}
{E_{\pi}\dd \sigma^{dAu}_{\pi}(\textrm{no nuclear effects})/ \dd ^3 p} \,\,\, .
\end{equation}

\section{Results on $dAu$ collisions  at $\eta=0$ in different centrality bins}

In Fig. 1 we display the measured $R_{dAu}^{\pi}$ for $\pi^0$ production at 
${\sqrt{s}_{NN}}=200$ GeV at different centralities:  
 $ 0-20\% $, $ 20-40\% $, $ 40-60\% $ and $ 60-88\% $~\cite{pre_phenix}. 
The first three  centrality bins (up to 60 \%) are similar to
each other, only a slight decrease can be seen.

\vspace*{-1.0truecm}
\begin{figure}[ht]
\label{fig1}
\begin{center}                                                              
\resizebox{160mm}{100mm}{\includegraphics{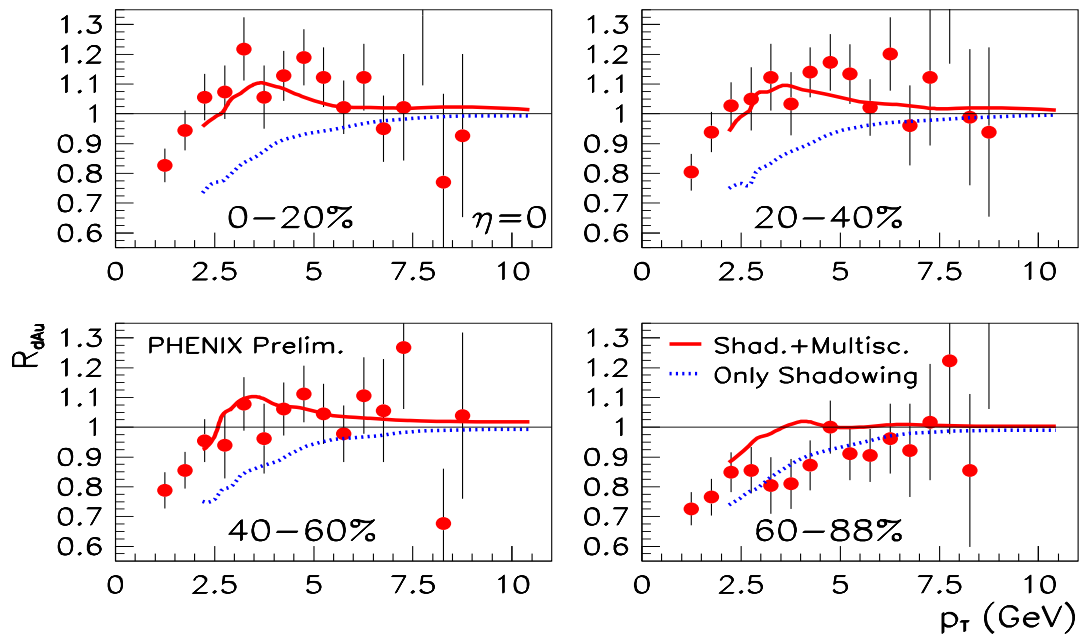}}
\caption{Theoretical results on $R_{dAu}^{\pi}$ at $\eta =0$, 
compared to PHENIX data on $\pi^0$~\cite{pre_phenix}.}
\end{center}                                                                
\end{figure}                                                               
Our theoretical pQCD results on $\pi^0$, containing multiscattering and
shadowing ({\it solid lines}), show similar tendency
and  overlap with the preliminary PHENIX data. 
We predicted this behavior in Ref.~\cite{dAu}. These data
indicate the plateau of the applied Woods\,--\,Saxon density profile
of the $Au$ nucleus and the applicability of a $b$-independent
shadowing in the model.
The difference between data and theory in the most peripheral case
requires further study.
To indicate the importance of nuclear multiscattering in $dAu$ collisions
we also show results containing only shadowing ({\it dotted lines}).

\section{Minimum bias results in $dAu$ collisions at forward rapidities  }

Figure 2 summarizes BRAHMS data on charged-hadron production
at forward pseudorapidities~\cite{pre_brahms},
CGC result at $\eta=3.2$ in the low-$p_T$ region~\cite{CGC} 
({\it dashed line}),
and our pQCD results on $\pi^0$ at $p_T > 2$ GeV.
We concentrate on the high-$p_T$ region, where no CGC result was available.
The increasing strength of conventional nuclear shadowing with increasing
pseudorapidity
is indicated by
pQCD results containing only this shadowing ({\it dotted lines}, $C=0$).
Including nuclear multiscattering (Cronin effect) we obtain
reasonable agreement between data and our pQCD calculations
on $R_{dAu}^{\pi}$ ({\it solid lines}). Our results indicate the validity
of pQCD description for $dAu$ \mbox{collisions at high $\eta$.}


\vspace*{-1.0truecm}
\begin{figure}[ht]
\label{fig2}
\begin{center}                                                              
\resizebox{140mm}{100mm}{\includegraphics{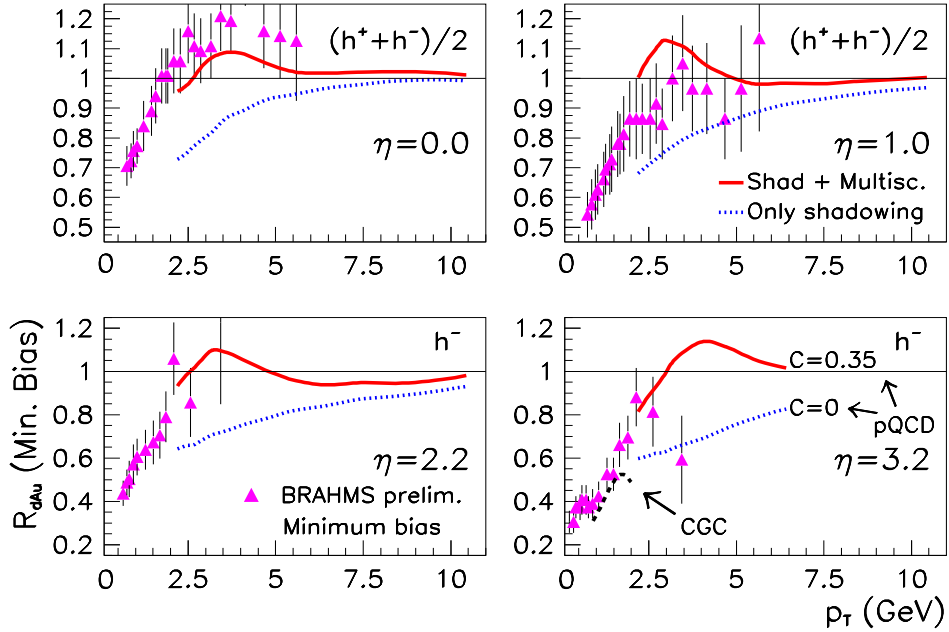}}
\caption{Theoretical pQCD results on $R_{dAu}^{\pi}$ 
for $\pi^0$ at $\eta =0$, $1$, $2.2$ and $3.2$,
compared to BRAHMS data on charged particles~\cite{pre_brahms}.
The CGC result is from Ref.~\cite{CGC}.}
\end{center}                                                                
\end{figure}

\section*{Acknowledgements}
One of the authors (GGB) would like to thank the organizers 
for local support. This work was supported by  
grants: T043455, T047050, and DE-FG02-86ER40251.


\end{document}